# Chapter 2
# Game-Theoretic Analysis of Cyber Deception: Evidence-Based Strategies and Dynamic Risk Mitigation

Tao Zhang[1], Linan Huang[1], Jeffrey Pawlick[1], and Quanyan Zhu[1]

[1]New York University, 2 Metrotech Center, Brooklyn, 11201, USA.

## 2.1 Abstract

Deception is a technique to mislead human or computer systems by manipulating beliefs and information. For the applications of cyber deception, non-cooperative games become a natural choice of models to capture the adversarial interactions between the players, and quantitatively characterizes the conflicting incentives and strategic responses. In this chapter, we provide an overview of deception games in three different environments and extend the baseline signaling game models to include evidence through side-channel knowledge acquisition to capture the information asymmetry, dynamics, and strategic behaviors of deception. We analyze the deception in binary information space based on signaling game framework with a detector that gives off probabilistic evidence of the deception when the sender acts deceptively. We then focus on a class of continuous one-dimensional information space and take into account the cost of deception in the signaling game. We finally explore the multi-stage incomplete-information Bayesian game model for defensive deception for advanced persistent threats (APTs). We use the perfect Bayesian Nash equilibrium (PBNE) as the solution concept for the deception games and analyze the strategic equilibrium behaviors for both the deceivers and the deceivees.

## 2.2 Introduction

Deception is a technique used to cause animals [9], human [13,35] or computer systems [3] to have false beliefs. The purpose of deception is to mislead the deceivees to behave against their interests but favorably to the deceiver. It is a fundamental type of interactions that can be found in applications ranging from biology [9] to criminology [35] and from economics [13] to the Internet of Things (IoT) [31]. Cyberspace creates particular opportunities for deception, since information lacks permanence, imputing responsibility is difficult [20], and some agents lack repeated interactions [24]. For instance, online interactions are vulnerable to identify theft [14] and spear phishing [1], and authentication in the IoT suffers from a lack of infrastructure and local computational resources [2].

Deception can be used as an approach for attacks. For example, phishing is a typical deception-based attack that is one of the top threat vectors for cyberattacks [32]. Phishing can be email-based in which a phisher manipulates the email to appear as a legitimate request for sensitive information [4, 5]. It can also be website-based in which the deceiver uses genuine looking content to camouflage a legitimate website to attract target deceivees to reveal their personal data such as credit card information and social security number. Defenders can also implement deception. Defenders in the security and privacy domains have proposed, *e.g.*, honeynets [7], moving target defense [39], obfuscation [30], and mix networks [36]. Using these techniques, defenders can obscure valuable data such as personally identifiable information or the configuration of a network. Using these approaches, defenders can send false information to attackers to waste their resources or distract them from critical assets. They can also obscure valuable data such as sensitive information or the configuration of a network to avoid direct accesses from the attackers. Both malicious and defensive deception have innumerable implications for cybersecurity. Successful deception fundamentally depends on the information asymmetry between the deceiver and the deceivee. Deceivees make indirect observations of the true state and then make decisions. Deceivers can take advantage of this by pretending to be a trustworthy information provider. It is possible to fool, mislead, or confuse the deceivees. But the deceivers need to plan their strategies and take actions that may be costly. Therefore, successful deception also requires the deceivers to have the ability to acquire information, accurately understand the goals of the deceivees, and make the induced actions predictable.

The deceivers strategically manipulate the private information to suit their own self-interests. The manipulated information is then revealed to the deceivees, who, on the other hand, make decisions based on the information received. It is important for the deceivee to form correct beliefs based on past observations, take into account the potential damage caused by deception, and strategically use the observed information for decision-making. If deception is necessary to achieve the deceivers' goal that would cause damages to the deceivees, the deceivees can then be prepared to invest resources in detecting and denying the deceptions as well as recovering the damage.

Modeling deceptive interactions online and in the IoT would allow government policymakers, technological entrepreneurs, and vendors of cyber-insurance to predict changes in these interactions for the purpose of legislation, development of new technology, or risk mitigation. Game-theoretic models are natural frameworks to capture the adversarial and defensive interactions between players [11, 16, 22, 23, 37, 42, 50, 51, 52, 53]. It can provide a quantitative measure of the quality of protection with the concept of Nash equilibrium where both defender and an attacker seek optimal strategies, and no one has an incentive to deviate unilaterally from their equilibrium strategies despite their conflict for security objectives. The equilibrium concept also provides a quantitative prediction of the security outcomes of the scenario the game model captures. With the quantitative measures of security, game theory makes security manageable beyond the strong qualitative assurances of cryptographic protections. Recently, we have seen game-theoretic methods applied to deal with problems in cross-layer cyber-physical security [23, 31, 40, 52, 54, 55], cyber deception [16, 27, 28, 29, 37, 53], moving target defense [39, 56, 57], critical infrastructure protection [19, 50, 51, 58, 59, 60, 61], adversarial machine learning [30, 62, 63, 64, 65], insider threats [66, 67], and cyber risk management [68, 69, 70, 71].

This chapter shows a class of modeling of deception based on signaling games to provide a generic, quantitative, and systematic understanding of deceptions in the cyber-domain. We show three variants of the model to illustrate the applications in different situations. We consider the cases when the deceivee is allowed to acquire knowledge through investigations or by deploying detectors. The baseline signaling game model is extended to include evidence through side-channel knowledge acquisition. We also show a multi-stage Bayesian game with two-sided incomplete information and present a dynamic belief update and an iterative decision process that are used to develop long-term optimal defensive policies to deter the deceivers and mitigate the loss.

### 2.2.1 Related Work

Deception game is related to a class of security games of incomplete information. For example, Powell in [34] has considered a game between an attacker and a defender, where the defender has private information about the vulnerability of their targets under protection. Powell models the information asymmetric interactions between players by a signaling game, and finds a pooling equilibrium where the defender chooses to pool, i.e., allocate resources in the same way for all targets of different vulnerabilities, and the attacker cannot know the true level of vulnerability of all targets. Brown et al. [6] have studied a zero-sum game between an attacker and a defender in the scenario of ballistic missile positioning. They have introduced the incomplete information to investigate the value of secrecy by restricting the players' access to information.

Previous literature has also considered deception in a variety of scenarios including proactive defense against advanced persistent threats [11,16,17,19,37], moving target defense [8, 39, 41], and social engineering [28, 29, 42]. Horák et al. [16] have considered a class of cyber deception techniques in the field of network security and studied the impact of the deception on attacker's beliefs using the quantitative framework of the game theory by taking into account the sequential nature of the attack and investigating how attacker's belief evolves and influences the actions of the players. Zhang et al., [37] have proposed an equilibrium approach to analyze the GPS spoofing in a model of signaling game with continuous type space. They have found a PBNE with pooling in low types and separating in high types, and provided an equilibrium analysis of spoofing. The hypothesis testing game framework in [43] has studied the influence of deceptive information on the decision making and analyzed the worst-case scenario by constructing equilibrium strategies. The model proposed by Ettinger et al. [10] has used an equilibrium approach to belief deception in bargaining problems when the agents only have coarse information about their opponent's strategy.

### 2.2.2 Organization of the Chapter

The chapter is organized as follows. In Section 2.2, we briefly describe common game-theoretic approaches for security models and introduce the basic signaling game model and define the perfect Bayesian Nash equilibrium (PBNE). In Section 2.3 we formulate the deception using a signaling game model with a detector over a binary information space, and describe the equilibrium result. In Section 2.4, we present a signaling-game-based framework of a deception game to model the strategic behaviors over a continuous one-dimensional information space. We also consider the knowledge acquisition for the receiver through investigations. In Section 2.5, we show a multi-stage Bayesian game framework to model the deception in advanced persistent threats. Section 2.6 discusses the results and provides concluding remarks.

## 2.3 Game Theory in Security

Game theory is the systems science that studies interactions between rational and strategic players (or agents) that are coupled in their decision makings. Players are rational in the sense that they choose actions to optimize their own objectives (e.g., utility or cost), which capture varying interaction contexts. Being strategic in game theory refers to that players choose their own actions by anticipating the actions of the other agents. Their decision makings are coupled because their objective functions depend both on their own actions, and on the actions of the other players. Among the game-theoretic cybersecurity models, Stackelberg game, Nash game, and signaling game account for the most commonly used approaches [28].

Stackelberg games consist of a *leader* ($L$) and a *follower* ($F$). $L$ has actions $a_L \in A_L$ and receives utility $U_L$, and $F$ has actions $a_F \in A_F$ and receives utility $U_F$. Once both players have taken actions, $L$ receives utility $U_L(a_L, a_F)$ and $U_F(a_L, a_F)$. In Stackelberg games, $F$ acts after knowing $L$'s action. Therefore, defensive cybersecurity models often take the defender as the leader and the attacker as the follower by considering the worst-case scenario that the attacker will observe and react to defensive strategies. Let $\mathcal{P}(A)$ denote the power set of the set $A$. Let $BR_F: A_L \to \mathcal{P}(A_F)$ denote the best response of $F$ to $L$'s action such that $BR_F(a_L)$ gives the optimal $a_F$ to respond to $a_L$. Best response can be one single action or a set of equally optimal actions. The function $BR_F$ is defined as

$$BR_F(a_L) = \underset{a_F \in A_F}{\mathrm{argmax}}\, U_F(a_L, a_F).$$

By anticipating $BR_F(a_L)$, $L$ chooses optimal action $a_L^*$ which satisfies

$$a_L^* \in \underset{a_F \in A_F}{argmax}\, U_L(a_L, BR_F(a_L)).$$

The action profile $(a_L^*, a_F^*)$ characterizes the equilibrium of the Stackelberg game, where $a_F^* \in BR_F(a_L^*)$.

In Nash games, on the other hand, players commit to his or her own strategy and move simultaneously or before knowing the other player's action [28]. Let $H$ and $T$ denote two players in a Nash game with actions $a_H \in A_H$ and $a_T \in A_T$, respectively. Define $BR_H(a_T)$ as the best response for $H$ that optimally respond to $T$'s action $a_T$. Similarly, let $BR_T(a_H)$ be the best response for $T$. *Nash equilibrium* is the solution concept of such games, which is defined by a strategy profile $(a_H^*, a_T^*)$, where

$$a_H^* \in BR_H(a_T),$$

$$a_T^* \in BR_T(a_H).$$

In other words, Nash equilibrium requires each player to simultaneously choose a strategy that is optimal given the other player's optimal strategy. In a *pure-strategy* Nash equilibrium, each player chooses one specific strategy (i.e., one pure strategy), while in a mixed-strategy equilibrium, at least on player randomizes over some or all pure strategies.

A signaling game is a two-player dynamic game of incomplete information. Signaling game typically names two players as sender (S, she) and receiver (R, he) [12]. Signaling game is information asymmetric because the sender privately possesses some information that is unknown to the receiver. The private information is usually referred to as state (or type) of the world. The sender communicates the receiver by sending a message, and the receiver only learns about the state through the message. Generally, the degree of conflict of interest between S and R may range from perfectly aligned (e.g., Lewis signaling game [21]) to completely opposite (e.g., zero-sum game [33]). The timing of the game is described as follows:

1. Nature randomly draws a state with a prior common to both players, and the sender privately observes the state.
2. The sender sends a message to the receiver.
3. The receiver takes an action upon observing the message.

One key feature of cyber deception is the multi-stage execution of the attack. A deceiver has to engage the deceivee in multiple rounds of interactions to gain the trust. The dynamic interactions have been observed in APT threats and the operation of honey devices [19]. Hence signaling games provide a suitable description of essential features of the deception, and the basic signaling game models will be elaborated in the following section.

### 2.3.1 Signaling Game Model

Let $\theta \in \Theta$ denote the state privately possessed by $S$ that is unknown to $R$. The state space $\Theta$ can be discrete (e.g., $\Theta_D \equiv \{\theta_0, \theta_1\}$) or continuous (e.g., $\Theta_C \equiv [\underline{\theta}, \overline{\theta}]$). For example, the state could represent, whether the sender is a malicious or benign actor, whether she has one set of preferences over another, samples of data stream, and location coordinate. For simplicity but without loss of generality, we focus on discrete state space in this introductory description of the signaling game. The state $\theta$ is drawn according to a prior distribution common to both players. Harsanyi conceptualized the state selection as a randomized move by a non-strategic player called *nature*. Let $p$ denote the probability mass function of the state, where $\sum_{\theta \in \Theta_D} p(\theta) = 1$. All aspects of the game except the value of the true state $\theta$ are common knowledge.

After privately observing the state $\theta$, $S$ chooses a message $m \in M$. Let $\sigma^S \in \Gamma^S$ denote the behavioral strategy of $S$ to choose $m$ based on $\theta$. She could use pure strategy $\sigma^S(\theta)$ as well as mixed strategy $\sigma^S(m|\theta)$, such that $\sigma^S(\theta)$ chooses message $m$ given $\theta$ and $\sigma^S(m|\theta)$ gives probability with which $S$ sends message $m$ given the state $\theta$. We assume the pure strategy $\sigma^S(\theta)$ induces a conditional probability $q^S(m|\theta) \in [0,1]$.

After receiving $m$, $R$ forms a posterior belief $\mu^R$ of the true state such that $\mu^R(\theta|m): \Theta \to [0,1]$ gives the likelihood with which R believes that the true state is θ given the message $m$. Based on the belief $\mu^R$, $R$ then chooses an action a $\in$ A according to a strategy $\sigma^R \in \Gamma^R$. Similarly, $R$ may employ pure strategy $\sigma^R(m)$ or mixed strategy $\sigma^R(a|m)$, where $\sigma^R(m)$ yields the action $R$ acts upon the message m, and $\sigma^R(a|m)$ produces the probability with which $R$ takes action $a$ given message m. The action a is the final decision of $R$ that represents the inference about the true state. Let $U^S$: Θ×M×A → $\mathbb{R}$ denote a utility function for $S$ such that $U^S(\theta, m, a)$ yields the utility of the player when her type is θ, she sends message $m$, and $R$ plays action $a$. Similarly, let $U^R$: Θ×M×A denote R's utility function so that $U^R(\theta, m, a)$ gives his payoff under the same scenario.

### 2.3.2 Perfect Bayesian Nash Equilibrium

In two player games, Nash equilibrium defines a strategy profile in which each player best responds to the optimal strategies of the other player. Signaling games motivate the extension of Nash equilibrium in two ways. First, information asymmetry requires $R$ to maximize his expected utility over the possible types of $S$. An equilibrium in which $S$ and $R$ best respond to each other's strategies given some belief $\mu^R$ is called a *Bayesian Nash equilibrium*. Furthermore, $R$ is required to update $\mu^R$ rationally. *Perfect Bayesian Nash equilibrium* (PBNE) captures this constraint and is described at Definition 1.

**Definition 2.1**: *(Perfect Bayesian Nash Equilibrium) A perfect Bayesian Nash equilibrium of a signaling profile $(\sigma^{S*}, \sigma^{R*})$ and a posterior belief system $\mu^R$ such that:*

1. $(C_1)$: $\forall \theta$, $\sigma^{S*} \in argmax_{\sigma^S} U^S(\theta, \sigma^S, \sigma^{R*})$,

2. $(C_2)$: $\forall m \in M$, $\sigma^{R*} \in argmax_{\sigma^R} \sum_\theta \mu^R(\theta|m) U^R(m, \sigma^R, \theta)$,

*and*

3. $(C_3)$: $\mu^R(\theta|m) = \frac{p(\theta)\sigma^{S*}(m|\theta)}{\sum_\theta p(\theta')\sigma^{S*}(m|\theta')}$, *if* $\sum_{\theta'} p(\theta')\sigma^{S*}(m|\theta') > 0$; *and* $\mu^R(\theta|m)$ *is any probability distribution on $\Theta$ if $\sum_{\theta'} p(\theta')\sigma^{S*}(m|\theta') = 0$.*

In Definition 2.1, $C_1$ and $C_2$ are the perfection conditions for $S$ and $R$, respectively, that characterizes the *sequential rationality* of both players. Specifically, $C_1$ captures that $S$ optimally determines $\sigma^{S*}$ by taking into account the effect of $\sigma^S$ on $\sigma^R$. $C_2$ says that $R$ responds rationally to $S$'s strategy given his posterior belief about the state $\theta$. $C_3$ states that the posterior belief is updated based on Bayes' rule. If the observation is a probability-0 event, then Bayes' rule is not applicable. In this case, any posterior beliefs over the state space is admissible. $C_3$ also implies that the consistency between the posterior beliefs and strategies: belief is updated depending on the strategy that is optimal given the belief. There are three categories of strategies: *separating*, *pooling*, and *partially-pooling* equilibria, which are defined based on the strategy of $S$. In separating PBNE (S-PBNE), $S$ chooses different strategies for different states. In this case, $R$ is able to identify each state with certainty. In pooling PBNE (P-PBNE), $S$ chooses the same strategy for different states. This strategy makes the corresponding message $m$ uninformative to $R$. In partially-pooling PBNE (PP-PBNE), however, $S$ chooses messages with different, but not completely distinguishable, strategies for different states. This makes the belief of $R$ remain uncertain.

**2.4 Binary State Space: Leaky Deception using Signaling Game with Evidence**

In [27], Pawlick et al. have modeled the strategic interactions between the deceiver $S$ and the deceivee $R$ over a binary information space by extending signaling games by including a detector that gives off probabilistic warnings called *evidence* when $S$ acts deceptively.

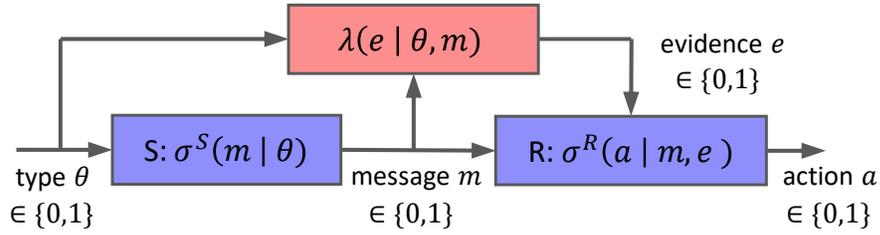

Fig. 2.1: Signaling games with evidence add the red detector block to the *S* and *R* blocks. The probability $\lambda(e|\theta, m)$ of emitting evidence *e* depends on *S*'s type $\theta$ and the message *m* that she transmits [27].

### 2.4.1 Game-Theoretic Model

With reference to Fig. 2.1, the detector emits evidences based on whether the message m is equal to the state $\theta$. The detector emits $e \in E \equiv \{0,1\}$ by the probability $\lambda(e \mid \theta, m)$. Let $e = 1$ denote an alarm and $e = 0$ no alarm. The evidence e is assumed to be emitted with an exogeneous probability that neither *R* nor *S* can control. In this respect, the detector can be seen as a second move by nature. Let $\beta \in [0,1]$ be the true-positive rate of the detector. For simplicity, both true-positive rates are set to be equal: $\beta = \lambda(1 \mid 0,1) = \lambda(1 \mid 1,0)$. Similarly, let $\alpha \in [0,1]$ denote the false-positive rate of the detector with $\alpha = \lambda(1 \mid 0,0) = \lambda(1 \mid 1,1)$. A valid detector has $\beta \geq \alpha$. This is without loss of generality, because otherwise $\alpha$ and $\beta$ can be relabeled. The timing of the game becomes:

1. Nature randomly draws state $\theta \in \{0,1\}$ according to p($\theta$).

2. *S* privately observes $\theta$ and then chooses a message m based on strategy $\sigma^S(m|\theta)$.

3. The detector emits evidence $e \in \{0,1\}$ with $\lambda(e|\theta, m)$.

4. After receiving both $m$ and $e$, *R* forms a belief system $\mu^R(\theta|m, e)$ and then chooses an action $a \in \{0,1\}$ according to strategy $\sigma^R(a|m, e)$.

The following assumptions characterizes a *cheap-talk signaling game with evidence.* The message $m$ is payoff-irrelevant in a cheap-talk signaling game.

**Assumption 2.1:** The utilities $U^S$ and $U^R$ satisfy the following assumptions:

1. $U^S$ and $U^R$ do not depend exogenously on $m$.

2. $\forall m, \tilde{m} \in M, U^R(0, m, 0) > U^R(0, \tilde{m}, 1)$.

3. $\forall m, \tilde{m} \in M, U^R(1, m, 0) < U^R(1, \tilde{m}, 1)$.

4. $\forall m, \tilde{m} \in M, U^S(0, m, 0) < U^S(0, \tilde{m}, 1)$.

5. $\forall m, \tilde{m} \in M, u^S(1, m, 0) > u^S(1, \tilde{m}, 1)$.

Assumption 2.1-1 implies that the interaction is a cheap-talk signaling game. Assumption 2.1-2 and 2.1-3 state that $R$ receives higher utility if he correctly chooses $a = \theta$ than if he chooses $a \neq \theta$. Finally, Assumption 2.1-4 and 2.1-5 say that $S$ receives higher utility if $R$ chooses $a \neq \theta$ than if he chooses $a = \theta$.

**Utilities.** Define an expected utility function $\overline{U}^S : \Gamma \times \Gamma^R \to \mathbb{R}$ such that $\overline{U}(\sigma^S, \sigma^R | \theta)$ gives the expected utility to $S$ when she plays strategy $\sigma^S$ given that the state is $\theta$. This expected utility is given by

$$\overline{U}^S(\sigma^S, \sigma^R | \theta) = \sum_{a \in A} \sum_{e \in E} \sum_{m \in M} \sigma^R(a | m, e) \lambda(e | \theta, m) \sigma^S(m | \theta) U^S(\theta, m, a). \quad (2.1)$$

The involvement of evidence $e$ in Eq. (2.1) is due to the dependence of $R$'s strategy $\sigma^R(a|m, e)$. $S$ must anticipate the probability of leaking evidence $e$ by using $\lambda(e|\theta, m)$. Similarly, define $\overline{U}^R : \Gamma \to \mathbb{R}$ such that $\overline{U}^R(\sigma^R | \theta, m, e)$ gives the expected utility to $R$ when he plays strategy $\sigma^R$ given message $m$, evidence $e$, and state $\theta$. The expected utility function is given by

$$\overline{U}^R(\sigma^R | \theta, m, e) = \sum_{a \in A} \sigma^R(a | m, e) U^R(\theta, m, a).$$

### 2.4.2 Equilibrium Concept

The involvement of the evidence extends the PBNE in Definition 2.2 as follows.

**Definition 2.2:** *(PBNE with Evidence) A PBNE of a cheap-talk signaling game with evidence is a strategy profile $(\sigma^{S*}, \sigma^{R*})$ and posterior beliefs $\mu^R(\theta | m, e)$ such that*

$$\forall \theta \in \Theta, \ \sigma^{S*} \in \underset{\sigma^S \in \Gamma^S}{\operatorname{argmax}} \overline{U}^S(\sigma^S, \sigma^{R*} | \theta), \quad (2.2)$$

$\forall m \in M, \forall e \in E,$

$$\sigma^{R*} \in \underset{\sigma^R \in \Gamma^R}{\arg\max} \sum_{\theta \in \Theta} \mu^R(\theta \mid m, e)\overline{U}^R(\sigma^R \mid \theta, m, e), \quad (2.3)$$

and if $\sum_{\tilde{\theta} \in \Theta} \lambda(e \mid \tilde{\theta}, m)\sigma^S(m \mid \tilde{\theta})p(\tilde{\theta}) > 0$, then

$$\mu^R(\theta \mid m, e) = \frac{\lambda(e \mid \theta, m)\mu^R(\theta \mid m)}{\sum_{\tilde{\theta} \in \Theta} \lambda(e \mid \tilde{\theta}, m)\mu^R(\tilde{\theta} \mid m)}, \quad (2.4)$$

where

$$\mu^R(\theta \mid m) = \frac{\sigma^S(m \mid \theta)p(\theta)}{\sum_{\hat{\theta} \in \Theta} \sigma^S(m \mid \hat{\theta})p(\hat{\theta})}. \quad (2.5)$$

If $\sum_{\tilde{\theta} \in \Theta} \lambda(e \mid \tilde{\theta}, m)\sigma^S(m \mid \tilde{\theta})p(\tilde{\theta}) = 0$, then $\mu^R(\theta \mid m, e)$ may be set to any probability distribution over $\Theta$.

Eq. (2.4)-(2.5) extend $C_3$ in Definition 2.2. First, $R$ updates her belief according to $m$ using Eq. (2.5); then $R$ updates her belief according to $e$ using Eq. (2.4). When $S$ plays pooling strategy, i.e., $\forall m \in M$, $\sigma^S(m|0) = \sigma^S(m|1)$. In this case, the message $m$ is uninformative, and $R$ updates his belief only depends on the evidence $e$, i.e.,

$$\mu^R(\theta \mid m, e) = \frac{\lambda(e \mid \theta, m)p(\theta)}{\sum_{\tilde{\theta} \in \Theta} \lambda(e \mid \tilde{\theta}, m)p(\tilde{\theta})}. \quad (2.6)$$

### 2.4.3 Equilibrium Results

Under Assumption 2.1-1 to 2.1-5, the cheap talk signaling game with evidence admits no separating PBNE. This results from the opposing utility functions of S and R. S wants to deceive R, and R wants to correctly guess the type. It is not incentive-compatible for S to fully reveal the type by choosing a separating strategy. For brevity, define the following notations:

$$\Delta_0^R \triangleq u^R(\theta = 0, m, a = 0) - u^R(\theta = 0, m, a = 1),$$

$$\Delta_1^R \triangleq u^R(\theta = 1, m, a = 1) - u^R(\theta = 1, m, a = 0).$$

$\Delta_0^R$ gives the benefit to R for correctly guessing the type when $\theta = 0$, and $\Delta_1^R$ gives the benefit to R for correctly guessing the type when $\theta = 1$. Lemmas 2.1-2.2 solve for $\sigma^{R*}$ within five regimes of the prior probability $p(\theta)$ of each type $\theta \in \{0,1\}$.

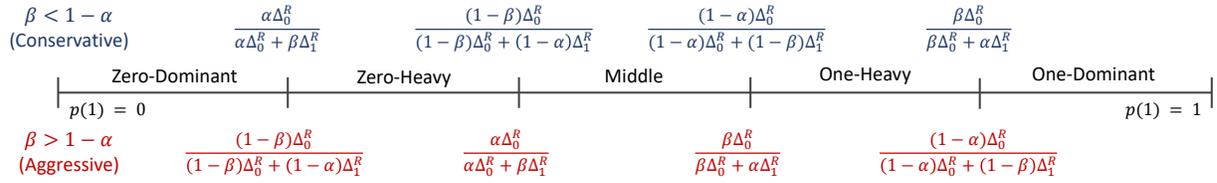

Fig. 2.2: PBNE differ within five prior probability regimes. In the Zero-Dominant regime, $p(\theta = 1) \approx 0$ i.e., type 0 dominates. In the Zero-Heavy regime, $p(\theta = 1)$ is slightly higher, but still low. In the Middle regime, the types are mixed almost evenly. The One-Heavy regime has a higher $p(\theta = 1)$, and the One-Dominant regime has $p(\theta = 1) \approx 1$. The definitions of the regime boundaries depend on whether the detector is conservative or aggressive.

**Lemma 2.1:** *For pooling PBNE, R's optimal actions $\sigma^{R*}$ for evidence e and messages m on the equilibrium path[1] vary within five regimes of $p(\theta)$. The top half of Fig. 2.2 lists the boundaries of these regimes for detectors in which $\beta < 1 - \alpha$, and the bottom half of Fig. 2.2 lists the boundaries of these regimes for detectors in which $\beta > 1 - \alpha$.*

The regimes in Fig. 2.2 shift towards the right as $\Delta_0^R$ increases. Intuitively, a higher $p(1)$ is necessary to balance out the benefit to R for correctly identifying a type $\theta = 0$ as $\Delta_0^R$ increases. The regimes shift towards the left as $\Delta_1^R$ increases for the opposite reason.

Lemma 2.2 gives the optimal strategies of R in response to pooling behavior within each of the five parameter regimes.

**Lemma 2.2:** *For each regime, $\sigma^{R*}$ on the equilibrium path is listed in Table 2.1 if $\beta < 1 - \alpha$ and in Table 2.2 if $\beta > 1 - \alpha$. The row labels correspond to the Zero-Dominant (O-D), Zero-Heavy (0-H), Middle, One-Heavy (1-H), and One-Dominant (1-D) regimes.*

Table 2.1: $\sigma^{R*}(1 \mid m, e)$ in Pooling PBNE with $\beta > 1 - \alpha$.

|  | $\sigma^{R*}(1 \mid 0,0)$ | $\sigma^{R*}(1 \mid 0,1)$ | $\sigma^{R*}(1 \mid 1,0)$ | $\sigma^{R*}(1 \mid 1,1)$ |
|---|---|---|---|---|
| 0-D | 0 | 0 | 0 | 0 |

---

[1] In pooling PBNE, the message "on the equilibrium path" is the one that is sent by both types of S. Messages "off the equilibrium path" are never sent in equilibrium, although determining the actions that R *would play* if S were to transmit a message off the path is necessary in order to determine the existence of equilibria.

| | | | | |
|---|---|---|---|---|
| 0-H | 0 | 1 | 0 | 0 |
| Middle | 0 | 1 | 1 | 0 |
| 1-H | 1 | 1 | 1 | 0 |
| 1-D | 1 | 1 | 1 | 1 |

Table 2.2: $\sigma^{R*}(1\,|m,e)$ in Pooling PBNE with $\beta > 1 - \alpha$.

| | $\sigma^{R*}(1\,|\,0,0)$ | $\sigma^{R*}(1\,|\,0,1)$ | $\sigma^{R*}(1\,|\,1,0)$ | $\sigma^{R*}(1\,|\,1,1)$ |
|---|---|---|---|---|
| 0-D | 0 | 0 | 0 | 0 |
| 0-H | 0 | 0 | 1 | 0 |
| Middle | 0 | 1 | 1 | 0 |
| 1-H | 0 | 1 | 1 | 1 |
| 1-D | 1 | 1 | 1 | 1 |

Lemmas 2.3-2.4 give conditions under which the beliefs $\mu^R$ exist such that each pooling strategy is optimal for both types of $S$.

**Lemma 2.3:** *Let $m$ be the message on the equilibrium path. If $\sigma^{R*}(1\,|\,m,0) = \sigma^{R*}(1\,|\,m,1)$, then there exists a $\mu^R$ such that pooling on message $m$ is optimal for both types of $S$. For brevity, let $a^* \triangleq \sigma^{R*}(1\,|\,m,0) = \sigma^{R*}(1\,|\,m,1)$. Then $\mu^R$ is given by,*

$$\forall e \in E,\ \mu^R(\theta = a^*\,|\,1-m, e) \geq \frac{\Delta^R_{1-a^*}}{\Delta^R_{1-a^*} + \Delta^R_{a^*}}.$$

**Lemma 2.4:** *If $\sigma^{R*}(1\,|\,m,0) = 1 - \sigma^{R*}(1\,|\,m,1)$ and $\beta \neq 1 - \alpha$, then there does not exist a $\mu^R$ such that pooling on message $m$ is optimal for both types of $S$.*

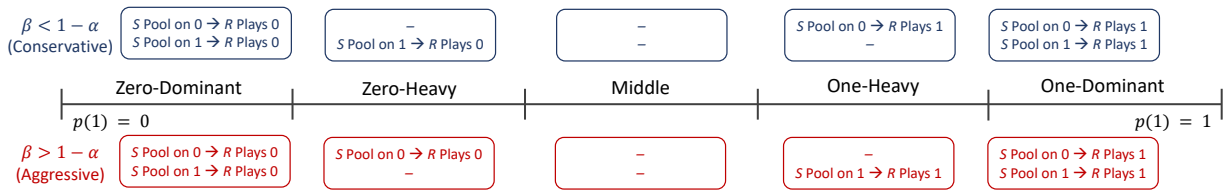

Fig. 2.3: PBNE in each of the parameter regions defined in Fig. 2.2. For $m \in \{0,1\}$, $S$ Pool on $m$ "Pool on $m$" ach $\sigma^{S*}(m|0) = \sigma^{S*}(m|1) = 1$. For $a \in \{0,1\}$, $R$ plays $a$ "plays $n$ ach $\sigma^{R*}(a|0,0) = \sigma^{R*}(a|0,1) = \sigma^{R*}(a|1,0) = \sigma^{R*}(a|1,1) = 1$.

Lemma 2.3 gives $\mu^{R*}$. The Dominant regimes support pooling PBNE on both messages. The Heavy regimes support pooling PBNE on only one message. The Middle regime does not support any pooling PBNE.

The pooling PBNE of the cheap-talk signaling game with evidence are summarized by Fig. 2.3. For $\beta \neq 1 - \alpha$, the Middle regime does not admit any pooling PBNE. This result is not found in conventional signaling games for deception, in which all regimes support pooling PBNE. It occurs because R's responses to message m depends on e, i.e., $\sigma^{R*}(1 \mid 0,0) = 1 - \sigma^{R*}(1 \mid 0,1)$ and $\sigma^{R*}(1 \mid 1,0) = 1 - \sigma^{R*}(1 \mid 1,1)$. One of the types of S prefers to deviate to the message off the equilibrium path. Intuitively, for a conservative detector, S with type $\theta = m$ prefers to deviate to message $1 - m$, because his deception is unlikely to be detected. On the other hand, for an aggressive detector, S with type $\theta = 1 - m$ prefers to deviate to message $1 - m$, because his honesty is likely to produce a false-positive alarm, which will lead R to guess $a = m$.

For $\beta \neq 1 - \alpha$, since the Middle regime does not support pooling PBNE, we search for partially-separating PBNE. In these PBNE, $S$ and $R$ play mixed strategies. In mixed-strategy equilibria in general, each player chooses a mixed strategy that makes the other players indifferent between the actions that they play with positive probability. Theorems 2.1-2.2 give the results.

**Theorem 2.1:** *(Partially-Separating PBNE for Conservative Detectors) For $\beta < 1 - \alpha$, within the Middle Regime, there exists an equilibrium in which the sender strategies are*

$$\sigma^{S*}(m = 1 \mid \theta = 0) = \frac{\beta^2}{\beta^2 - \alpha^2} - \frac{\alpha \beta \Delta_1^R}{(\beta^2 - \alpha^2)\Delta_0^R}\left(\frac{p(1)}{1-p(1)}\right),$$

$$\sigma^{S*}(m = 1 \mid \theta = 1) = \frac{\alpha \beta \Delta_0^R}{(\beta^2 - \alpha^2)\Delta_1^R}\left(\frac{1-p(1)}{p(1)}\right) - \frac{\alpha^2}{\beta^2 - \alpha^2},$$

*the receiver strategies are*

$$\sigma^{R*}(a = 1 \mid m = 0, e = 0) = \frac{1 - \alpha - \beta}{2 - \alpha - \beta},$$
$$\sigma^{R*}(a = 1 \mid m = 0, e = 1) = 1,$$
$$\sigma^{R*}(a = 1 \mid m = 1, e = 0) = \frac{1}{2 - a - b},$$
$$\sigma^{R*}(a = 1 \mid m = 1, e = 1) = 0,$$

*and the beliefs are computed by Bayes' Law in all cases.*

**Theorem 2.2:** *(Partially-Separating PBNE for Aggressive Detectors) For any $g \in [0,1]$, let $\bar{g} \triangleq$*

$1 - g$. For $\beta > 1 - \alpha$, within the Middle Regime, there exists an equilibrium in which the sender strategies are

$$\sigma^{S*}(m = 1 \mid \theta = 0) = \frac{\overline{\alpha}\overline{\beta}\Delta_1^R}{\left(\overline{\alpha}^2 - \overline{\beta}^2\right)\Delta_0^R}\left(\frac{p(1)}{1-p(1)}\right) - \frac{\overline{\beta}^2}{\overline{\alpha}^2 - \overline{\beta}^2},$$

$$\sigma^{S*}(m = 1 \mid \theta = 1) = \frac{\overline{\alpha}^2}{\overline{\alpha}^2 - \overline{\beta}^2} - \frac{\overline{\alpha}\overline{\beta}\Delta_0^R}{\left(\overline{\alpha}^2 - \overline{\beta}^2\right)\Delta_1^R}\left(\frac{1-p(1)}{p(1)}\right),$$

the receiver strategies are

$$\sigma^{R*}(a = 1 \mid m = 0, e = 0) = 0,$$
$$\sigma^{R*}(a = 1 \mid m = 0, e = 1) = \frac{1}{\alpha + \beta},$$
$$\sigma^{R*}(a = 1 \mid m = 1, e = 0) = 1,$$
$$\sigma^{R*}(a = 1 \mid m = 1, e = 1) = \frac{\alpha + \beta - 1}{\alpha + \beta},$$

and the beliefs are computed by Bayes' Law in all cases.

In Theorem 2.1, $S$ chooses the $\sigma^{S*}$ that makes $R$ indifferent between $a = 0$ and $a = 1$ when he observes the pairs $(m = 0, e = 0)$ and $(m = 1, e = 0)$. This allows $R$ to choose mixed strategies for $\sigma^{R*}(1 \mid 0,0)$ and $\sigma^{R*}(1 \mid 1,0)$. Similarly, $R$ chooses $\sigma^{R*}(1 \mid 0,0)$ and $\sigma^{R*}(1 \mid 1,0)$ that make both types of $S$ indifferent between sending $m = 0$ and $m = 1$. This allows $S$ to choose mixed strategies. A similar pattern follows in Theorem 2.2 for $\sigma^{S*}$, $\sigma^{R*}(1 \mid 0,1)$, and $\sigma^{R*}(1 \mid 1,1)$.

## 2.5 Continuous State Space: Knowledge Acquisition and Fundamental Limits of Deception

In [38], Zhang et al. proposes a game-theoretic framework of a deception game to model the strategic behaviors of the deceiver $S$ and the deceivee $R$ and construct strategies for both attacks and defenses over a continuous one-dimensional state space. $R$ is allowed to acquire probabilistic evidence about the deception through investigations, and misrepresenting the state is costly for $S$. The deceivability of the deception game is analyzed by characterizing the PBNE.

### 2.5.1 Game-Theoretic Model

We assume that the state $\theta$ is continuously distributed over $\Theta \equiv [\underline{\theta}, \overline{\theta}]$ according to a differentiable probability distribution $F(\theta)$, with strictly positive density $f(\theta)$ for all $\theta \in \Theta$. Again, all aspects of the game except the value of the true state $\theta$ are common knowledge.

**Message and Report.** In this game model, we use the message to describe the format of information about the state S communicates to R. We introduce a notion report to represent the value of state carried by the message. After privately observing the state $\theta$, S first determines a report $r \in \Theta$ for the true state $\theta$, and then sends R a message $m \in M$, where M is a Borel space of messages. Let $\Omega: M \to \Theta$ denote the report interpretation function such that $\Omega(m)$ gives the report r carried in $m$. Given the true state $\theta$, we say m tells the truth if $\Omega(m) = \theta$. We assume that for each state $\theta \in \Theta$, there is a sufficiently large number of messages that yields the same report, and each $m \in M$ has a unique value of report $\Omega(m)$. In other words, the message space can be partitioned as $M = \cup_r M_r$, with $|M_r| \to \infty$ for all r and $M_r \cap M_{r'} = \emptyset$ if $r \neq r'$, and $\forall m \in M_r$, $\Omega(m) = r$. This assumption can capture the feature of rich language in practical deceptions. We further assume that message $m$ is formed by "common language" that can be understood precisely by both S and R. In other words, function $\Omega$ is commonly known by both players.

**Strategies and actions.** Let $\sigma^S: \Theta \to \Theta$ be the strategy of S such that $r = \sigma^S(\theta)$ determines the report r of the true state $\theta$. Let $\eta^S: \Theta \times \Theta \to M$ be the message strategy of S associated with $\sigma^S$ such that $m = \eta^S(r)$ selects the message $m$ from $M_r$ when the strategy $\sigma^S(\theta)$ determines the report r and the true state is $\theta$. Given $\theta$, the strategy $\sigma^S(\theta)$ determines the set of messages $M_{\sigma^S(\theta)}$ for $\eta^S$ to choose from, and $\eta^S$ determines which specific message $m \in M_{\sigma^S(\theta)}$ to send. We assume $\sigma^S(\theta)$ associated with $\eta^S$ induces a conditional probability $q^S(m|\theta)$. After receiving m, R chooses an action $a \in A \equiv \Theta$ according to a strategy $\sigma^R: \Theta \times M \to A$ using $r = \Omega(m)$. $\sigma^R(r, m)$ gives the action R acts upon the message m (and thus $r = \Omega(m)$). The action a is the final decision of R that represents the inference about the true state.

**Utilities.** The utility functions of S is given by $U^S(a, \theta, r) \equiv U^A(a, \theta) - kU^D(r, \theta)$ where $U^A(a, \theta) \equiv -(a - (\theta + b))^2$ is the utility depending on the induced action a in R, $U^D \equiv -(r - \theta)^2$ is the utility related to the misrepresentation of the true state, and $k \geq 0$ quantifies the intensity of $U^D$. On the deceivee's side, his utility is given by $U^R \equiv -(a - \theta)^2$, which takes into account the risk induced by R's misinference of the true state $\theta$ via his action $a$. Define, for all $\theta \in \Theta$, $\alpha^S(\theta) \equiv \text{argmin}_a C^S(a, \theta, r)$, and $\alpha^R(\theta) \equiv \text{argmin}_a C^R(a, \theta)$; i.e., $\alpha^R(\theta)$ and $\alpha^S(\theta)$ are two actions taken by R as functions of $\theta$ that are the most preferred by R and S, respectively.

**Beliefs.** Based on m (and thus $r = \Omega(m)$) and his prior belief $f(\theta)$, R forms a posterior belief $\mu^R: \Theta \to [0,1]$ of the true state $\theta \in \Theta$. The posterior belief $\mu^R(\theta|m)$ gives the likelihood with which R believes that the true state is $\theta$ based on m. R then determines which action to choose based on his belief $\mu^R$.

### 2.5.2 Deceivability

We restrict attention to a class of monotone inflated deception, in which the strategy profile $(\sigma^S, \sigma^R)$ satisfies conditions in the following definition.

**Definition 2.3:** A deception with $S$'s strategy $\sigma^S$ and $R$'s belief $\mu^R$ is monotone if

1. $\sigma^S(\theta)$ is a non-decreasing function of $\theta$;
2. $\sigma^R(r, m)$ is a non-decreasing function of $r$.

The deceivability can be quantified as follows.

**Definition 2.4:** Given the state $\theta \in [\theta'', \theta']$, $S$'s strategy $\sigma^S(\theta) = r$, and message strategy $\eta^S(r) = m$,

- $R$ is *undeceivable* over $[\theta'', \theta']$ if $\sigma^R(r, m) = \alpha^R(\theta) \equiv \theta$, for all $\theta \in [\theta'', \theta']$. Here, $\sigma^S(\theta) \neq \sigma^S(\theta')$ for all $\theta \neq \theta' \in [\theta'', \theta']$, and $\eta^S(r) \in M_r$. The corresponding $\mu^R$ is *informative*. The interval $[\theta'', \theta']$ is called *undeceivable region* (UR).

- $R$ is *deceivable* over $[\theta'', \theta']$ if the only knowledge she has is that $\theta$ lies in $[\theta'', \theta']$. $R$ chooses $\sigma^R(r, m) = \hat{a}^R(\theta'', \theta')$, by maximize the expected utility over $[\theta'', \theta']$, i.e.,

$$\hat{a}^R(\theta'', \theta') \in \operatorname*{argmax}_{\sigma^R \in A} \int_{\theta'}^{\theta''} U^R(\sigma^R, \theta) f(\theta) d\theta.$$

Here, $\sigma^S(\theta)$ and $\eta^S(r)$, respectively, choose the same report $r$ and the same message $m$, for all $\theta \in [\theta'', \theta']$. Thus, given $m$, $q^S(m|\theta)$ is the same for all $\theta \in [\theta'', \theta']$, where $q \in (0,1)$. The corresponding $\mu^R$ is *uninformative*. The interval $[\theta'', \theta']$ is called *deceivable region* (DR).

### 2.5.3 Knowledge Acquisition: Evidence

We allow $R$ to acquire additional knowledge through investigations when the state is in a DR, $[\theta'', \theta']$, by partitioning it into multiple intervals, denoted by a strictly increasing sequence, $< \theta_0 = \theta'', \theta_1, \ldots, \theta_J = \theta' >$. Then $R$ conducts investigations for each interval. Here, we consider the case when there are two investigation intervals to simplify the analysis. Let $\sigma^c \in (\theta'', \theta')$ be the *investigation partition state* such that $[\theta'', \theta']$ is partitioned into two non-overlapping investigation regions $\Theta^0 = [\theta'', \theta^c]$ and $\Theta^1 = [\theta^c, \theta']$. Let $\Psi \in \Gamma = \{\Psi^0, \Psi^1\}$, where $\Psi^i$ denote the event $\{\theta \in \Theta^i\}$, for $i = 0, 1$, with the probability $P(\Psi^i) = \frac{\int_{\Theta^i} f(\tilde{\theta}) d\tilde{\theta}}{\int_{[\theta'', \theta']} f(\tilde{\theta}) d\tilde{\theta}}$. The investigation for $\Theta^0$ and $\Theta^1$ generates noisy evidence $e \in E = \{0,1\}$, where $e = i$ represents $\Psi^i$, for $i = 0, 1$.

Suppose that the investigation emits evidence by the probability $\gamma(e|\Psi, m)$. Let $x = \gamma(e = 0|\Psi^0, m)$ and $y = \gamma(e = 1|\Psi^1, m)$ be the two true positive rates, which are private information of $R$.

With a slight abuse of notation, let $\sigma^R(\Psi, m, e): \Gamma \times M \times E \to A$ be the strategy of $R$ with evidence $e$. Fig. 2.4 depicts the signaling game model for the deception with knowledge acquisition through investigation.

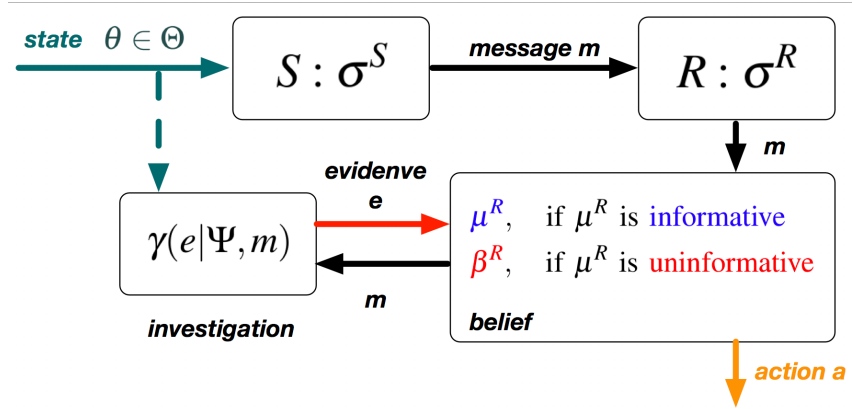

Fig. 2.4: Signaling games with evidence acquisition by investigation. The probability $\gamma(e|\Psi, m)$ of emitting evidence $e$ depends on the event $\Psi$ and the message $m$ sent by $S$. If the belief $\mu^R$ is informative, $\mu^R$ is used; if $\mu^R$ is uninformative, $\beta^R$ is used as the posterior.

### 2.5.4 Equilibrium Concept

The knowledge acquisition of the signaling game over continuous state space extends the PBNE in Definition 1 to the following.

**Definition 2.5:** (Perfect Bayesian Nash Equilibrium) A PBNE of the game is a strategy profile $(\sigma^S, \sigma^R)$ and a posterior belief system $(\mu^R, \beta^R)$ that satisfy the following conditions:
- *(Deceiver's Sequential Rationality)* $S$ maximizes her expected utility given the deceivee's strategy $\sigma^R$ and the distribution of the evidence $e$: for each $\theta \in \Theta$,

$$\sigma^{S*}(\theta) \in \underset{\sigma^S}{\mathrm{argmax}}\, U^S(\sigma^{R*}, \theta, \sigma^S).$$

- *(Deceivee's Sequential Rationality)* $R$ maximizes his expected utility given $S$'s strategy $\sigma^{S*}$ and his posterior belief $\mu^R(\theta|m)$: for any $m \in M$,

  • if $\mu^R(\theta|m)$ is informative, $\sigma^{R*}(r, m) \in \underset{\sigma^R \in A}{\mathrm{argmax}} \int_{\theta \in \Theta} U^R(\sigma^R, \theta) \mu^R(\theta|m) d\theta$;

- if $\mu^R(\theta|m)$ is uninformative over $\Theta_U \equiv [\theta'', \theta'] \subseteq \Theta$, $\sigma^{R*}(\Psi, m, e) \in$
  $\text{argmax}_{\hat{a}^i} \sum_{i=0}^{1} \int_{\Theta_U^i} \beta(\Psi^i|m, e) U^R(\hat{a}^i, \tilde{\theta}) f(\tilde{\theta}) d\tilde{\theta}$,

  where $\theta_U^0 \equiv [\theta'', \theta^c]$, $\theta_U^1 \equiv [\theta'', \theta^c]$, and $\hat{a}^i \equiv \text{argmax} \int_{\Theta_U^i} U^R(a, \tilde{\theta}) d\tilde{\theta}$.

- *(Consistent Belief)* The posterior belief of $R$ is updated according to Bayes' rule (i.e., $\mu^R(\theta|m)$ is *informative*), as

$$\mu^R(\theta|m) = \frac{f(\theta) q^S(m|\theta)}{\int_\Theta f(\tilde{\theta}) q^S(m|\tilde{\theta}) d\tilde{\theta}}.$$

- If $\int_\Theta f(\tilde{\theta}) q^S(m|\tilde{\theta}) d\tilde{\theta} = 0$, $\mu^R(\theta|m)$ may be set to any probability distribution over $\Theta$. If $\mu^R$ is *uninformative*, i.e.,

$$\mu^R = \frac{f(\theta)}{\int_{[\theta'', \theta']} f(\tilde{\theta}) d\tilde{\theta}}.$$

- $R$ acquires evidence through investigation, and updates belief using evidence as,

$$\beta^R(\Psi|e, m) = \frac{\gamma(e|\Psi, m) P(\Psi)}{\sum_{j=0}^{1} \gamma(e|\Psi^j, m) P(\Psi^j)},$$

- and if $\sum_{j=0}^{1} \gamma(e|\Psi^j, m) P(\Psi^j) = 0$, $\beta^R(\Psi|e, m)$ may be set to any probability distribution over $\Theta$.

In separating equilibrium, the deceiver sends message $m$ with different values of report $\Omega(m)$ for different states. Separating equilibria are also called *revealing* equilibria because the strategic deceivee can infer the true state even if $\Omega(m)$ does not tell the truth. In pooling equilibrium, the deceiver sends message $m \in M_r$ with the same value of report $\Omega(m) = r$ for all states. In partial-pooling equilibrium, however, the deceiver sends the message with the same report for some states and different reports for other states. Clearly, the PBNE strategy $\sigma^{S*}$ associated with a DR (resp. UR) is pooling (resp. separating) strategy.

### 2.5.5 Equilibrium Results

From the definition of UR, the equilibrium strategy of $R$ gives the most preferred action, $\alpha^R(\theta) \equiv \theta$, for all $\theta$ in the UR. Therefore, in any differentiable S-PBNE, the utility $U^S$ and the strategy $\sigma^S$ have to satisfy the following first-order optimality condition given $\sigma^{R*}(\theta) = \alpha^R(\theta)$ according to the sequential rationality: $U_1^S(\alpha^R(\theta), \theta, \sigma^S(\theta)) \frac{d\alpha^R(\theta)}{d\theta} + U_3^S(\theta, \theta, \sigma^S(\theta)) \frac{d\sigma^S(\theta)}{d\theta} = 0$.

Lemma 2.5 summarizes the property of the strategy $\sigma^{S*}$ in any UR.

**Lemma 2.5:** If $[\theta_s, \theta_l]$ is an undeceivable region, then for each $\theta \in [\theta_s, \theta_l]$, the equilibrium strategy $\sigma^{S*}(\theta) > \theta$ and it is a unique solution of the differential equation

$$\frac{d\sigma^S(\theta)}{d\theta} = \frac{b}{k(\sigma^S(\theta) - \theta)},$$

with initial condition $\sigma^{S*}(\theta_s) = \theta_s$.

Lemma 2.5 underlies the following proposition.

**Proposition 2.1:** With initial condition $\sigma^{S*}(\underline{\theta}) = \underline{\theta}$, there exists a cut-off state $\hat{\theta} < \overline{\theta}$ such that a unique solution $\sigma^{S*}$ to the differential equation in Lemma 2.5 is well-defined on $[\underline{\theta}, \hat{\theta}]$ with $\sigma^{S*}(\hat{\theta}) = \overline{\theta}$, and there is no solution on $(\hat{\theta}, \overline{\theta}]$.

Proposition 2.1 notes that in S-PBNE, the optimal strategy $\sigma^{S*}$ of $S$ has to choose a report $r$ that is strictly larger than the true state $\theta$, but eventually $\sigma^{S*}$ runs out of such report for $\theta > \hat{\theta}$. Proposition 2.1 implies that there is no S-PBNE strategy of $S$ for all $\theta > \hat{\theta}$, because there are not enough states to support the monotone S-PBNE strategy of $S$ for the state in $(\hat{\theta}, \overline{\theta}]$. This suggests a class of PP-PBNE for the state space $\Theta$, which is separating in low states and pooling in higher states. For convention, let $\sigma^{S,p}: \Theta \to \Theta$ and $\eta^{S,p}$, respectively, denote the P-PBNE strategy and the associated message strategy of $S$. We define this class of PP-PBNE by introducing a boundary state in the following precise sense.

**Definition 2.6:** We say that the strategy $\sigma^S$ is a SLAPH (Separating in Low states And Pooling in High states) strategy if there exists a boundary state $\theta_B \in [\underline{\theta}, \hat{\theta}]$ such that
1. *(S-PBNE)* $\sigma^{S*}(\theta) = r$ with $\eta^{S*}(r) \in M_r$, for all $\theta \in [\underline{\theta}, \theta_B)$, and $\sigma^{S*}(\theta) \neq \sigma^{S*}(\theta')$ for all $\theta \neq \theta' \in [\underline{\theta}, \theta_B)$;

2. *(P-PBNE)* $\sigma^{S*,p}(\theta) = \overline{\theta}$ with $\eta^{S*,p}(\theta) \in M_{\overline{\theta}}$, for all $\theta \in [\theta_B, \overline{\theta}]$.

In any SLAPH equilibrium, both players have no incentive to deviate from the equilibrium strategies. This requires the boundary state $\theta_B$ to be consistent in the sense that the equilibrium at $\theta_B$ is well-defined. Specifically, the utility of $S$ has to satisfy the following *boundary consistency* (BC) condition at $\theta_B$:

$$U^S(\sigma^{R*}(\sigma^{S,p}(\theta_B), m_p), \theta_B, m_p) = U^S(\alpha^R(\theta_B), \theta_B, m_s),$$

where $m_p \in M_{\bar{\theta}}$ and $m_s \in M_{\sigma^{S*}(\theta_B)}$. The BC condition implies that $S$ is indifferent between sending $m_p \in M_{\bar{\theta}}$ with $a^* = \sigma^{R*}(\sigma^{S*}(\theta_B), m_p)$ and sending $m_s \in M_{\sigma^{S*}(\theta_B)}$ with action $a^* = \theta_B$.

The conflict of interest, b, is a utility-relevant parameter for S that can induce incentives for S to reveal partial information about any state $\theta \in [\theta_B, \bar{\theta}]$ to R while her utility-maximizing P-PBNE strategy $\sigma^{S*}$ is maintained. This can be achieved based on the assumption $|M_{\bar{\theta}}| \to \infty$ and the fact that $C^D$ is equally expensive for all the messages chosen for all state $\theta \in [\theta_B, \bar{\theta}]$. Specifically, the P-PBNE region $[\theta_B, \bar{\theta}]$ can be further partitioned into multiple pools. First, some notations for describing the multiple pools are needed. Let $\Theta^P \equiv (\theta_0, \theta_1, \ldots, \theta_{K-1}, \theta_K)$ be a partition of $[\theta_B, \bar{\theta}]$, with $\theta_0 = \theta_B < \theta_1 < \cdots < \theta_K = \bar{\theta}$. We call each interval $\Theta_{j,j+1} = [\theta_j, \theta_{j+1}]$ is a pool. With an abuse of notation, let $\eta^{S*,P}(\sigma^{S*}(\theta), \theta)$ denote the message strategy that chooses a message $m \in M_{\sigma^{S*}(\theta)}$ for a state $\theta$. In each pool $\Theta_{j,j+1}$, $\eta^{S*,P}(\bar{\theta}, \theta)$ chooses the same message $m \in M_{\bar{\theta}}$, for all $\theta \in \Theta_{j,j+1}$, $j = 0, \ldots, K-1$. After R determines a pool $\Theta_{j,j+1}$, she acquires evidence $e \in \{e_0, e_1\}$ through investigations by dividing $\Theta_{j,j+1}$ into two sub-intervals $\Theta^0_{j,j+1} \equiv [\theta_j, \theta^I_{j,j+1}]$ and $\Theta^1_{j,j+1} \equiv [\theta^I_{j,j+1}, \theta_{j+1}]$. Let $\Psi_{j,j+1} \in \Gamma_{j,j+1} \equiv \{\Psi^0_{j,j+1}, \Psi^1_{j,j+1}\}$ such that $\Psi^i_{j,j+1}$ represents the event $\{\theta \in \Theta^i_{j,j+1}\}$, with probability $P(\Psi^i_{j,j+1}) = \dfrac{\int_{\Theta^i_{j,j+1}} f(\tilde{\theta}) d\tilde{\theta}}{\int_{\Theta_{j,j+1}} f(\tilde{\theta}) d\tilde{\theta}}$, for $i = 0, 1$. On the equilibrium path, R must play $\sigma^{R*}(\Psi_{j,j+1}, m_j, e)$ as defined in Definition 2.5 for any $m_j$ such that $\eta^{S*,P}(\bar{\theta}, \theta) = m_j$ for all $\theta \in \Theta_{j,j+1}$. Define

$$\hat{a}^i(\theta_j, \theta_{j+1}) \equiv \underset{a}{argmax} \int_{\Theta^i_{j,j+1}} U^R(a, \theta) f(\theta) d\theta,$$

for $i = 0, 1$.

The necessary and sufficient conditions for the existence of SLAPH equilibrium are summarized in the following theorem.

**Theorem 2.3: (Necessary condition.)** In any SLAPH equilibrium, there exists a boundary state $\theta_B$ such that the pooling interval $[\theta_B, \bar{\theta}]$ can be partitioned into multiple pools denoted by a strictly increasing sequence $(\theta_0, \theta_1, \ldots, \theta_{K-1}, \theta_K)$ with $\theta_0 = \theta_B$ and $\theta_K = \bar{\theta}$, such that, for all $j = 0, \ldots, K-1$,

$$U^A(\bar{a}(\theta_j, \theta_{j+1}), \theta_{j+1}) = U^A(\bar{a}(\theta_{j+1}, \theta_{j+2}), \theta_{j+1}), \quad (2.7)$$

$$U^S(\bar{a}(\theta_0, \theta_1), \theta_B, \bar{\theta}) = U^S(\theta_B, \theta_B, \sigma^{S*}(\theta_B)), \quad \text{if } \theta_B > \underline{\theta}, \quad (2.8)$$

where $\bar{a}(\theta_j, \theta_{j+1}) = \sum_{i=0}^{1} P(\Psi^i) \hat{a}^i(\theta_j, \theta_{j+1})$, for all $j = 0, \ldots, K-1$. **(Sufficient Condition.)** Given the multiple-pool PBNE characterized by Eq. (2.7)-(2.8), and if $\theta_B = \underline{\theta}$ and

$$U^S\big(\overline{a}(\theta_0,\theta_1),\underline{\theta},\overline{\theta}\big) \leq U^S\big(\alpha^R(\underline{\theta}),\underline{\theta},\sigma^{S*}(\underline{\theta})\big), \quad (2.9)$$

there exists an SLAPH equilibrium.

Note that $U^D$ is equal for all $\theta \in [\theta_B, \overline{\theta}]$. Eq. (2.7) says that at each link state $\theta_{j+1}$ connecting $\Theta_{j,j+1}$ and $\Theta_{j+1,j+2}$, the utilities induced by $\overline{a}(\theta_j,\theta_{j+1})$ and $\overline{a}(\theta_{j+1},\theta_{j+2})$, respectively, should keep the same. Otherwise, $S$ has incentives to deviate from the current partition to combine these two consecutive pools by sending the message that induces more profitable $U^A$ but the same $U^D$. This is not ideal for $R$ because larger pools make the posterior less informative that could decrease the utilities for $R$. Similarly, Eq. (2.8) says that at the boundary state $\theta_B$, $S$ should be indifferent between playing S-PBNE strategy and inducing action $\sigma^{R*}(\theta_B)$ versus playing P-PBNE and introducing action $\overline{a}(\theta_0,\theta_1)$. Inequality (2.9) notes that if the boundary state $\theta_B = \underline{\theta}$, then $S$ is indifferent between pooling with $[\underline{\theta},\theta_1]$ and reporting $\overline{\theta}$ for $\underline{\theta}$ versus separating at $\underline{\theta}$. The existence of SLAPH requires (2.7)-(2.9) to be jointly satisfied.

## 2.6 Adaptive Strategic Cyber Defense for APT in Critical Infrastructure Network

### 2.6.1 Multi-Stage Dynamic Bayesian Game Model

In [18] and [19], Huang et al. have proposed a multi-stage Bayesian game framework to capture the stealthy, dynamic, and adaptive natures of the advanced persistent threats (APTs) as shown in Fig. 2.5.

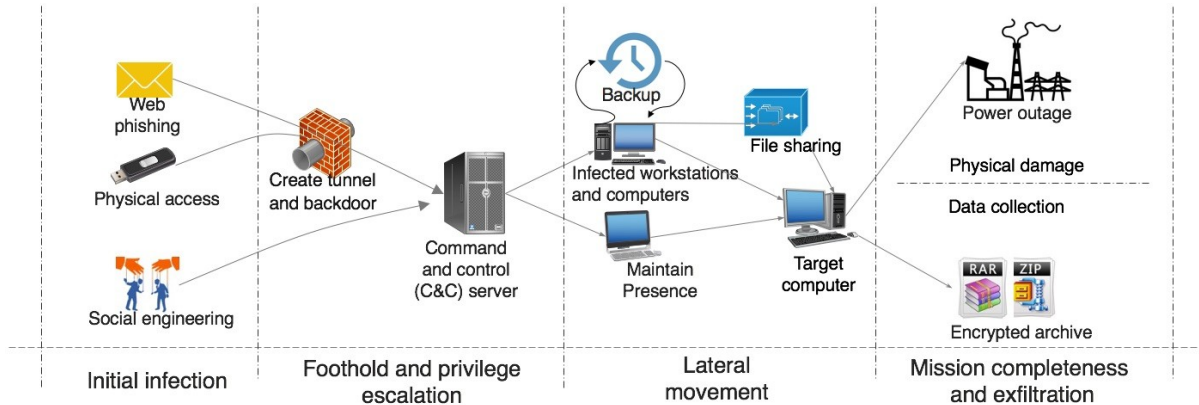

Fig. 2.5: The APTs' life cycle includes a sequence of phases and stages such as the initial entry, privilege escalations, and lateral movements. APTs use each stage as a stepping stone for the next and aim to cause physical damages or collect confidential data.

**Continuous Type:** Due to the cyber deception, the system defender $R$ cannot directly determine whether a user is legitimate or not even he can observe the user's apparent behaviors. Thus, user

$S$ has a type $\tilde{\theta}$ which is a random variable and the realization $\theta \in \Theta := [0,1]$ is private information of $S$. The value of the type indicates the strength of the user in terms of damages that she can inflict on the system. A user with a larger type value indicates a higher threat level. At each stage $k \in \{0,1,\cdots,K\}$, $S$ chooses an action $m_k \in M_k$ and $R$ chooses an action $a_k \in A_k$. The user's actions represent the apparent behaviors and observable activities from log files such as a privilege escalation request and sensor access. A defender cannot identify the user's type from observing his actions. The defender's action includes prevention and proactive behaviors such as restricting the escalation request or monitoring the sensor access. The action pair $(m_k, a_k)$ is known to both players after stage $k$ and forms a *history* $h_k := \{m_0, \cdots, m_{k-1}, a_0, \cdots, a_{k-1}\}$. The state $x_k \in \mathcal{X}_k$ shows the system status at each stage $k$ such as the location of the APTs. Since the initial state $x_0$ and history $h_k$ uniquely determine the state, $x_k$ contains information of history up to $k$ and has the transition kernel described by $x_{k+1} = g_k(x_k, m_k, a_k)$. The behavior mixed strategies $\sigma_k^S \in \Gamma_k^S : \mathcal{X}_k \times \Theta \to \Delta M_k$ and $\sigma_k^R \in \Gamma_k^R : \mathcal{X}_k \to \Delta A_k$.

**Believe Update:** To strategically gauge the user's type, the defender specifies a belief $\mu_k^R : \mathcal{X}_k \mapsto \Delta \Theta$ as a distribution over the type space according to the information available at stage $k$. The prior distribution of the user's type is known to be $f$ and the belief of the type updates according to the Bayesian rule.

$$\mu_{k+1}^R(\theta | x_{k+1}) = \frac{\mu_k^R(\theta | x_k) \sigma_k^R(a_k | x_k, \theta)}{\int_0^1 \mu_k^R(\hat{\theta} | x_k) \sigma_k^R(a_k | x_k, \hat{\theta}) d\hat{\theta}}.$$

**Utility Function**: The user's type influences $P_i$'s immediate payoff received at each stage $k$, i.e., $U_k^S : \mathcal{X}_k \times M_k \times A_k \times \Theta \mapsto \mathbb{R}$. For example, a legitimate user's access to the sensor benefits the system while a pernicious user's access can incur a considerable loss. Define $\sigma_{k':K}^S := \{\sigma_k^S \in \Gamma_k^S\}_{k=k',\cdots,K} \in \Gamma_{k':K}^S$ as a sequence of policies from $k'$ to $K$.

The defender has the objective to maximize the cumulative expected utility:

$$\bar{U}_{k':K}^R(\sigma_{k':K}^S, \sigma_{k':K}^R, x_{k'}) := \sum_{k=k'}^{K} E_{\theta \sim \mu_k^R, m_k \sim \Gamma_k^S, a_k \sim \Gamma_k^R} U_k^R(x_k, m_k, a_k, \theta),$$

and the user's objective function is

$$\bar{U}_{k':K}^S(\sigma_{k':K}^S, \sigma_{k':K}^R, x_{k'}, \theta) = \sum_{k=k'}^{K} \sum_{a_k \in A_k} \sigma_k^R(a_k | x_k) \sum_{m_k \in M_k} \sigma_k^S(m_k | x_k, \theta) U_k^S.$$

### 2.6.2 Equilibrium Concept

The insider threats of APTs lead to the following definition of perfect Bayesian Nash equilibrium (PBNE) where the defender chooses the most rewarding policy to confront the attacker's best-response policies.

**Definition 2.7:** In the two-person multi-stage game with a sequence of beliefs $\mu_k^R, k \in \{k', \cdots, K\}$ satisfying the Bayesian update in and the cumulative utility function $\overline{U}_{k':K}^S$, the set $R_2^{\theta, x_{k'}}(\sigma_{k':K}^R) :=$ $\{\gamma \in \Gamma_{k':K}^S : \overline{U}_{k':K}^S(\sigma_{k':K}^R, \gamma, x_{k'}, \theta) \geq \overline{U}_{k':K}^S(\sigma_{k':K}^R, \sigma_{k':K}^S, x_{k'}, \theta), \forall \Gamma_{k':K}^R \in \Gamma_{k':K}^R, \forall x_{k'} \in \mathcal{X}_{k'}, \theta \in \Theta\}$ is $S$'s **best-response set** to $R$'s policy $\sigma_{k':K}^R \in \Gamma_{k':K}^R$ under state $x_{k'}$ and type $\theta$.

**Definition 2.8:** In the two-person multi-stage Bayesian game with $R$ as the principal, the cumulative utility function $\overline{U}_{k':K}^S, \overline{U}_{k':K}^R$, the initial state $x_{k'} \in \mathcal{X}_{k'}$, the type $\theta \in \Theta$, and a sequence of beliefs in , a sequence of strategies $\sigma_{k':K}^{R*} \in \Gamma_{k':K}^R$ is called a perfect Bayesian Nash equilibrium (**PBNE**) for the principal, if

$$\overline{U}_{k':K}^{R*}(x_{k'}) := \inf_{\sigma_{k':K}^S \in R_2^{\theta, x_{k'}}(\sigma_{k':K}^{R*})} \overline{U}_{k':K}^{R*}(\sigma_{k':K}^{R*}, \sigma_{k':K}^S, x_{k'}).$$

A strategy $\sigma_{k':K}^{S*} \in argmax_{\sigma_{k':K}^S \in \Gamma_{k':K}^S} \overline{U}_{k':K}^S(\sigma_{k':K}^{R*}, \sigma_{k':K}^S, x_{k'}, \theta)$ is a PBNE for the agent $S$.

A conjugate-prior method allows online computation of the belief and reduces Bayesian update into an iterative parameter update. The forwardly updated parameters are assimilated into the backward dynamic programming computation to characterize a computationally tractable and time-consistent equilibrium solution based on the expanded state space.

### 2.6.3 Case Study

We consider a four-stage transition with the first three stages related to cyber transition of the APTs and the last stage related to the benchmark Tennessee Eastman (TE) process as the targeted physical plant. The TE process involves two irreversible reactions to produce two liquid (liq) products. The process shuts down when the safety constraints are violated such as a high reactor pressure, a high/low separator/stripper liquid level. The attacker can revise the sensor reading, trigger an undesired feedback-control, and cause a loss. The state $x_k$ has five possible values representing the working status of the system where the state 1 is most desirable and state 5 is the least desirable. We can obtain the normal operation reward and the reward of attacks from the simulation. The belief state $\{\alpha_k, \beta_k\}$ uniquely determines the belief distribution which is assumed to take the form of the beta distribution. The larger $\alpha_k$ means that the user is more likely to have a smaller type value and have lower threats to the system.

As shown in Fig. 2.6, a high value $\overline{U}_{k':K}^{R*}(x_{k'})$ for the defender is the result of a healthy system state $x_k$ as well as a belief of a low-threat user. At the most desirable system state $x_k = 1$, attackers will not attack because the reward incurred is insufficient. Then the defender does not need to defend and obtains the maximum utility.

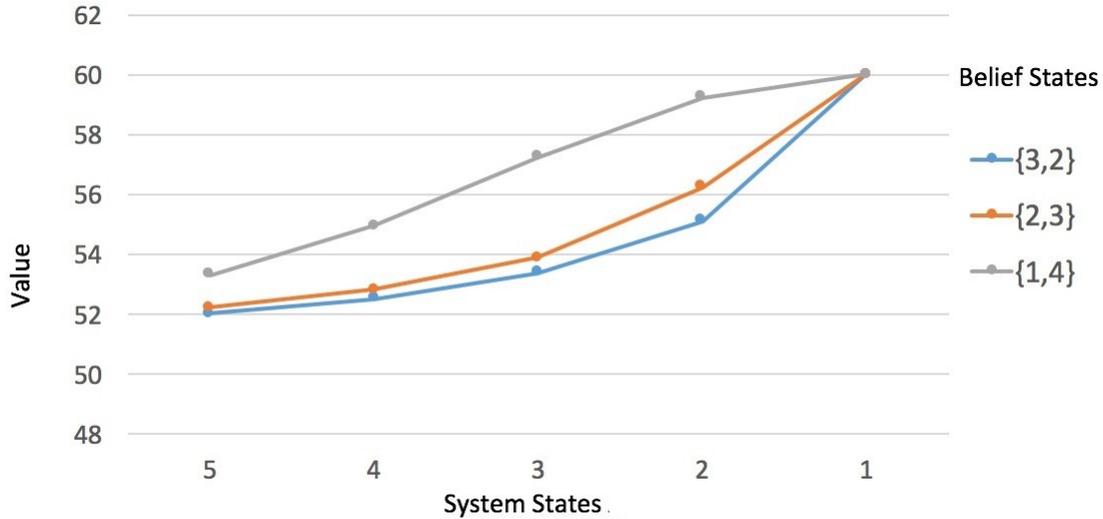

Fig. 2.6: Value function $\overline{U}_{k':K}^{R*}(x_{k'})$ under different expanded states $\{x_k, \alpha_k, \beta_k\}$.

To investigate the effect of the defender's belief, we change the belief state $(\alpha_k, \beta_k)$ from $(9,1)$ to $(1,9)$, which means that the defender grows optimistically that the user is of a low threat level with a high probability. Since players' value functions are of different scales in terms of the attack threshold and the probability, we normalize the value functions with respect to their maximum values to illustrate their trends and make them comparable to the threshold and the probability as shown in Fig. 2.7. When $\alpha_k$ is large, the defender chooses to protect the system with a high probability, which completely deters attackers with any type values because the probability to attack is 0.

As the defender trusts more about the user's legitimacy, the defending probability decreases to 0 when $\alpha_k = 1$. Since the defender is less likely to defend, the attacker bears a smaller threshold to launch the attack. The resulted defending policy captures a tradeoff between security and economy and guarantees a high value for defenders at most of the belief states.

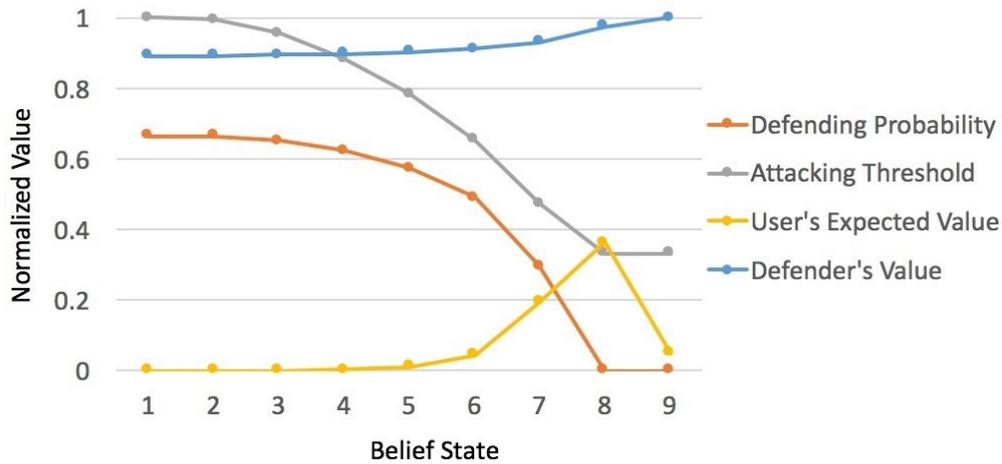

Fig. 2.7: The effect of the defender's belief.

The central insight from the multi-stage analysis is the adversary's tradeoff between the instantaneous reward and the hiding to arrive at a more favorable state in the future stages. The higher the belief of the defender in $S$ as a legitimate user, the less probability he will act defensively and thus the attacker has a smaller threshold to launch the attack.

## 2.7 Conclusion

Deception is a technique that can be viewed as an advanced approach to secure the devices or attacks. Understanding deception quantitatively is pivotal to provide rigor, predictability, and design principles. In this chapter, we have formulated signaling-game theoretic models of deceptions over discrete and continuous information spaces. We have studied leaky deception models and extended the baseline perfect Bayesian Nash equilibrium (PBNE) to versions involving knowledge acquisitions characterized by evidences. We have analyzed the impacts of evidence on the belief updates and strategy constructions on the equilibrium path.

In the binary state space, the leaky deception game with evidence admits an equilibrium that includes a regime in which the deceivee should choose whether to trust the deceiver based on the evidence, and regimes in which the deceivee should ignore the message and evidence and merely guess the private information based only on the prior probabilities. For the deceiver, the equilibrium results imply that it is optimal to partially reveal the private information in the former regime.

We have also studied leaky deception games over a continuous one-dimensional information space. We have studied the PBNE as the solution concept to analyze the outcome of the deception game and characterize the deceivability of the game. The proposed deception game admits a class of PBNE called SLAPH (Separating in Low states And Pooling in High states). The necessary and sufficient conditions for the existence of such PBNE are given. However, a full undeceivable region does not exist and there exists a deceivable region. We have also shown that the deceivable region can be partitioned into multiple sub-deceivable regions without decreasing total utilities for the deceiver when the conflict of interest is insignificant.

Furthermore, we have explored a multi-stage incomplete information Bayesian game model for defensive deception frameworks for critical infrastructure networks with the presence of advanced persistent threats (APT). With the multi-stage and multi-phase structure of APTs, the belief of the defender is formed dynamically using observable footprints. The conjugate priors are used to reduce Bayesian updates into parameter updates, which leads to a computationally tractable extended-state dynamic programming that admits an equilibrium solution consistent with the forward belief update and backward induction. Tennessee Eastman process has been used as a case study to demonstrate the multi-stage deception game. The numerical simulations have shown that the game-theoretic defense strategies have significantly improved the security of the critical infrastructures.